# Tailoring the breathing-mode distortions in nickelate/ferroelectric heterostructures

Guillaume Krieger[1,⊥], Chia-Ping Su[2], Hoshang Sahib[1], Raymond Fan[3], Paul Steadman[3], Alexandre Gloter[2], Nathalie Viart[1] and Daniele Preziosi[*,1]

[1]Université de Strasbourg, CNRS, IPCMS UMR 7504, F-67034 Strasbourg, France
[2]Laboratoire de Physique des Solides, CNRS, Université Paris-Saclay, 91405 Orsay, France [3]Diamond Light Source, Diamond House, Harwell Science and Innovation Campus, Didcot, Oxfordshire OX11 0DE, United Kingdom

In transition metal oxides electron-electron interaction and lattice degree of freedom are basic ingredients of emergent phenomena, such as metal-to-insulator transition (MIT) and superconductivity. Perovskite rare-earth nickelates are largely studied for their temperature-driven MIT which is accompanied by a breathing mode distortion, and associated to a bond-disproportionation of the expanded ($3d^8L^0$) and compressed ($3d^8L^2$) $NiO_6$ octahedra. Steric effects control the onset temperature of the MIT, the latter being concomitant or not with a complex antiferromagnetic spin arrangement depending upon the choice of the rare earth ion ($T_{MIT} \geq T_{Néel}$). Interface engineering of oxygen octahedra tilting, as imposed by the symmetry and orientation of the substrate, has resulted in an efficient pathway to modify both $T_{MIT}$ and $T_{Néel}$, hence, suggesting a key role of the electron-phonon coupling for both transport and magnetic properties in nickelate thin films. Here, via a combination of resonant elastic X-ray scattering and transport experiments, we show a control over both $T_{MIT}$ and $T_{Néel}$ in heteroepitaxial $PbZr_{0.2}Ti_{0.8}O_3$(d)/$NdNiO_3$(7 nm)//$SrTiO_3$ heterostructures, which are characterized by different strains and polarization states of the $PbZr_{0.2}Ti_{0.8}O_3$ layer grown at different thicknesses d. We found the expected $NdNiO_3$ bulk behaviour ($T_{MIT} = T_{Néel}$), for a fully relaxed $PbZr_{0.2}Ti_{0.8}O_3$ layer showing a monodomain polarization state. On the other side, an almost 30 K difference ($T_{MIT} > T_{Néel}$), is found for a fully strained $PbZr_{0.2}Ti_{0.8}O_3$ characterized by a multidomain texture of the polarization state. We discuss our results in terms of an altered breathing distortion pattern of the underlying nickelate layer as supported by X-ray absorption spectroscopy measurements. We infer that locally different polar distortions controlled by a combination of polarization direction and strength of the strain state play the main role in the observed $T_{MIT}$ and $T_{Néel}$ variations.

## 1. Introduction

The building of complex heterostructures with transition metal oxides offers the opportunity to control their different phases through strain and lower dimensionality. This has potential implications in the designing of novel functional devices[1]. Rare-earth perovskite nickelates (*R*$NiO_3$, *R* being a rare-earth), possess a rich phase diagram where a temperature dependent metal-to-insulator transition (MIT) is accompanied or followed by a complex antiferromagnetic (AFM) ordering[2–4] (R≠La), whose onset temperatures ($T_{MIT}$ and $T_{Néel}$, respectively) are controlled by the distortion of the unit cell. The anti-ferromagnetic (AFM) order occurs at the Bragg vector $q_{AFM}$ = (¼,¼,¼)$_{pc}$ in the pseudocubic unit cell, as early proved by neutron diffraction experiments on powder samples[5]. More recent resonant elastic x-ray scattering (REXS) measurements gave information also about the staggered ordering of the spins at the two inequivalent Ni sites below $T_{Néel}$ (collinear vs. non-collinear)[6,7]. Depending upon the choice of *R*, the Ni-O-Ni bond angle deviating from the ideal value of 180°, imposes a variation of the Ni 3d-O 2p orbital's overlap, and thus of the bandwidth of the material[8]. The MIT is accompanied by a structural transition from the orthorhombic *Pbmn* towards the monoclinic $P2_1/n$ space group, and this symmetry lowering takes place via a so-called breathing-mode distortion characterized by a particular bond-disproportionation. As a result, in the insulating phase, the negative charge transfer nickelates[9] present inequivalent $NiO_6$ octahedra which alternate with long (L) and short (S) Ni-O bonds and $3d^8\underline{L}^0$ and $3d^8\underline{L}^2$ configurations, respectively ($\underline{L}$ being a ligand hole)[3]. Double cluster simulations have shown that this bond-disproportionation are at the base of the peculiar Ni-$L_3$ features observed via X-ray absorption spectroscopy (XAS) measurements at low temperature, each of which, correspond to the expanded and compressed $NiO_6$ octahedra[10]. Additionally, the antiferromagnetic Ni-O coupling gives rise to different magnetic moment for each sublattice, and the Ni-spin tends toward 1 for the expanded octahedra (L), and to 0 for the compressed one (S). Those features at the Ni-$L_3$ edge can also be used as an indirect proxy of the resistive state as already reported by some of the authors[11]. Up to now, electron and/or hole doping[12], hydrostatic pressure[13] or even a control over the oxygen vacancies[14] have been used to modulate the properties of the nickelates and tailor their phase diagram in bulk-form, while epitaxial strain[15,16], substrate orientations[17] or even heterostructuring[18–20] are mostly used in the case of nickelates thin films. Here, we want to take advantage of the thickness-dependent epitaxial strain and polarization domain texture of a $PbZr_{0.2}Ti_{0.8}O_3$ (PZT) layer[21] grown onto a $NdNiO_3$ thin film with the primary goal to exploit the bandwidth-controlled phase diagram of the perovskite nickelates by providing a novel lever over the lattice[9,10]. It has been already demonstrated that the presence of a large amount of oxygen vacancies are strictly related to the nucleation of large area characterized by spurious phases, and usually indexed as Ruddsleden-Popper phases with predominant $Ni^{2+}$ oxidation state and consequent modification of both transport and magnetic properties[22]. On the other side, defect-free PZT thin films requires a fine tuning of the deposition conditions to preserve the ferroelectric properties at the smallest thickness[23]. In this respect, nickelates/ferroelectric heterostructures have been very rarely investigated so far[24–26], and it would be of relevance to understand how the growth of the domain structure of the PZT layer can affects the nickelates' functional properties. This would allow to make a first parallel with similar investigated manganite-based heterostructures[27–29].

We have grown $PbZr_{0.2}Ti_{0.8}O_3(d)/NdNiO_3(7\ nm)//SrTiO_3(001)$ (PZT/NNO//STO) heterostructures (d = 6, 13, 30 nm), as well as a bare NNO layer with a non-polar STO capping-layer of 6 nm, that hereafter will be referred as PZT-0. We investigated the influence of the PZT polarization domain structure and strain on the paramagnetic-to-antiferromagnetic and metal-to-insulator phase transitions of the underlying NNO layer. Unfortunately, we could not operate any remnant reversal of the PZT polarization in macroscopic switching experiments when a NNO layer was used as bottom electrode due to a large imprint. We performed temperature dependent REXS measurements, which are largely used to probe the AFM ordering in nickelate thin films, to unambiguously determine the value of $T_{Néel}$ by tracking the intensity of the observed peak in a cooling process[30]. For further details please refer to the dedicate section in Methods. As a result, transport together with REXS measurements were used to observe at which extent both $T_{Néel}$ and $T_{MIT}$ varied as a function of the PZT thickness, respectively. Scanning transmission electron microscopy (STEM) cross-section images were used to visualize domains growth of the PZT layer as a function of its thickness, and X-ray absorption spectroscopy (XAS) measurements have been performed at both Ni $L_{3,2}$-edges and O K-edge. By considering the overall PZT strain state in connection with the observed polar domain structure and the XAS properties, we discuss our transport and REXS data in terms of a modified bond-disproportionation pattern of the NNO films. Specifically, the combined substrate-induced strain and PZT thickness values control the polar domain texture showing an as-grown upwards/downwards texture-orientation, and the latter via intrinsic polar distortions[29] may alters the Ni-O-Ni bond angles with different magnitude, as depending upon the PZT strain state itself. Fully strained PZT layers are found to exert a larger influence on the bond-disproportionation state of the NNO most likely due to a larger interface polar-distortion effect ($T_{MIT} > T_{Néel}$). On the other side, for fully relaxed PZT layers characterized by a monodomain polar state the polar-distortions are found to be less effective on the alteration of the Ni-O-Ni bond angles, since both standard XAS features, and bulk-like transport/magnetic behaviour ($T_{MIT}=T_{Néel}$) are fully recovered.

# Methods

PZT(d)/NNO heterostructures were grown by pulsed laser deposition (PLD) onto STO (001) substrates. The growth was monitored *in situ* by reflection high energy electron diffraction (RHEED). The structural characterization of the samples was obtained by X-ray diffraction (XRD) with a Rigaku Smartlab diffractometer equipped with a rotating anode and a monochromatic copper radiation ($\lambda$ = 0.154056 nm), in both symmetric (θ-2θ scans) and asymmetric (reciprocal space mappings) modes. Resistivity versus temperature [ρ(T)] measurements were performed in a four-points Van der Pauw geometry by applying a current of 10 μA in a Dynacool system (Quantum Design). Scanning transmission electron microscopy (STEM) was performed on a Cs-corrected NION STEM system operated at a 100 keV voltage. Cross sectional samples have been obtained by focussed ion beam polishing technique. Resonant elastic X-ray scattering (REXS) and X-ray absorption spectroscopy (XAS) measurements were performed at the I10 beamline of the Diamond synchrotron in Oxfordshire (London-England). This is a soft X-ray beamline delivering photons with linear or circular polarisations in the 400-1600 eV energy range. XAS measurements were performed at the Ni $L_{3,2}$-edges in both total electron yield (TEY) and fluorescence yield (FY) modes. Setting the energy at the Ni $L_3$-edge allows optimising the sensitivity of the scattering measurements to magnetism. At this resonance energy a Ni electron with a well-defined spin state is transferred from the 2p to the 3d states, and probes the magnetic state of the material. The intensity of the scattering is in this case very sensitive to the polarisation of the incident beam. We performed the REXS experiments at RASOR end station[23] where two main axes, θ and 2θ, rotate the sample relative to the incident beam and detector around the sample's centre of rotation, respectively. The (¼,¼,¼)$_{pc}$ reflection was measured by rocking the θ sample angle through the diffraction condition at the photon energy corresponding at the first Ni-$L_3$ feature (ca. 853 eV). A Janis ST400 cryostat enabled the cooling and temperature control. Since the spin configuration is described as a stacking of ferromagnetically ordered planes perpendicular to the pseudo-cubic [111]$_{pc}$ crystallographic direction the samples were mounted on a copper wedge structure with an tilt-angle ($\Psi$, *cf.* to Fig. 6a) of 45°. Thanks to the good versatility of the I-10 beam-line at Diamond we could operate a fine tuning of the Y-angle in such a way to orient the [111]$_{pc}$ crystallographic direction (55⁰ from the [001]$_{pc}$) within the scattering plane.

# Growth and characterization of the heterostructures

Table 1 summarizes the parameters used for the growth of the studied PZT(d)/NNO(7nm)//STO heterostructures (HTs), based on parameters previously optimized by some of the authors and discussed in Refs. [28,31]. The HTs will be referred to as PZT-6, PZT-13 and PZT-30 for each chosen PZT thickness value of 6, 13 and 30 nm, respectively. To prevent the presence of oxygen vacancies, after growth, all the heterostructures underwent an *in-situ* thermal annealing process consisting in maintaining each grown sample in the PLD chamber at a temperature of 500 °C in an $O_2$ pressure of 100 mbar for five hours, followed by a cooling down process in the same oxygen atmosphere.

Table 1 Growth parameters of the PZT/NNO//STO(001) heterostructures.

| Target | Fluence | Temperature | Laser Frequency | Deposition atmosphere | Post-treatment |
|---|---|---|---|---|---|
| PbZr$_{0.2}$Ti$_{0.8}$O$_3$ | 2.0 J/cm² | 600°C | 10 Hz | 0.3 mbar O$_2$ | -O$_2$ annealing: 100 mbar, 500°C, 5 h -Cooling down: 5°C/min, 100 mbar O$_2$ |
| NdNiO$_3$ | 3.8 J/cm² | 675°C | 2 Hz | | |

Through Reflection High Energy Electron Diffraction (RHEED) we studied the growth dynamics of the PZT layer onto the NNO while gathering important information about the PZT/NNO interface itself. Figure 1 shows a sequence of RHEED patterns acquired during the growth of the HTs. The NNO layer shows a fully 2D RHEED pattern indicating a layer-by-layer growth process. However, at the very beginning of the PZT growth, the relatively large repetition rate used, set in an island growth mode as demonstrated by a fully 3D RHEED pattern. The latter is progressively modified upon increasing the PZT thickness, until, for a threshold value of 30 nm a rather 2D-like RHEED pattern is recovered.

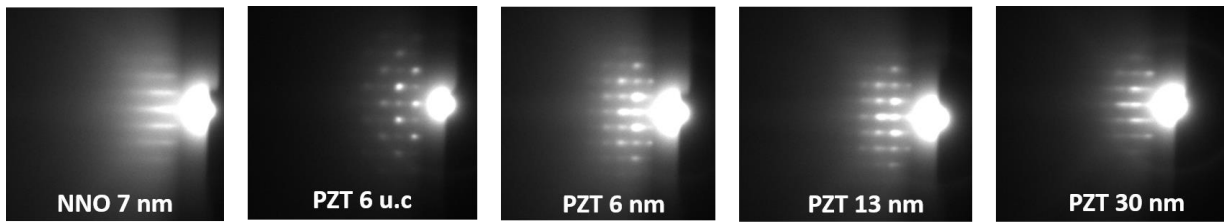

Figure 1 RHEED diffraction patterns of (from left to right) a 7 nm thick NNO layer, a 6 u.c PZT layer grown on top of the 7 nm NNO showing a 3D growth, the PZT-6 sample at the end of its growth, still 3D, the PZT-13 sample at the end of its growth which is also still 3D, and the PZT-30 sample which shows a 2D pattern at the end of its growth.

Figure 2 summarizes the structural characterization of all the HTs via X-ray diffraction measurements. In Figure 2a one can observe that the PZT peaks show a clear thickness dependence. The sample with the thinnest PZT layer, *i.e.* PZT-6, presents a rather broad feature at each expected reflection position. This may be due to the fact that the several islands evidenced by the 3D RHEED pattern develop slightly different tetragonal structural distortions and/or orientations. When the thickness of the PZT layer is increased, the corresponding diffraction peaks become better defined and more intense, which speaks in favor of a monodomain polar state. Moving to the structural properties of the NNO layer ($a_{pc}$ = 0.381 nm for the bulk), we can see from the zoom around the (002) peaks (Fig. 2b), that the latter move at higher values with increasing PZT thickness, and getting closer to the (002) NNO peak of the PZT-0 film. As a result, the NNO out-of-plane lattice parameter (c-axis) was found to unexpectedly depend upon the PZT layer thickness, as summarized in Figure 2c. The c-axis measured for the PZT-0 is the one expected for NNO grown onto STO and capped with a non-polar oxide[31]. The NNO c-axis value tends toward the expected one for larger PZT thickness values. To study at which degree the tensile strain imposed by the substrate comes into play, we performed reciprocal space mapping (RSM) measurements around the asymmetric STO (013) reflection for each HTs. We found that the PZT-6 layer is fully strained to the NNO (Fig. 2e), and this strain state is progressively relaxed for increasing thickness of the ferroelectric layer until the PZT-30 sample (Fig. 2g) exhibits the bulk-like values for the PZT lattice parameters, *i.e.* a = 0.395 nm and c = 0.414 nm. As already reported in literature in the case of NNO thin films under tensile strain, the presence of defects, off-stoichiometry and/or oxygen vacancies result in an increased lattice volume[11,25]. Here, since all the NNO layers are fully strained to the STO substrate, an expansion of the lattice volume can be uniquely obtained by an increase of the c-axis. The NNO c-axis is the largest for the thinnest PZT top-layer, and it decreases for increasing PZT thickness. Since the NNO layers were grown with the same parameters and underwent the same annealing processes, it is most likely that we need to call for different mechanisms than random defects and/or oxygen vacancies to explain this volume change.

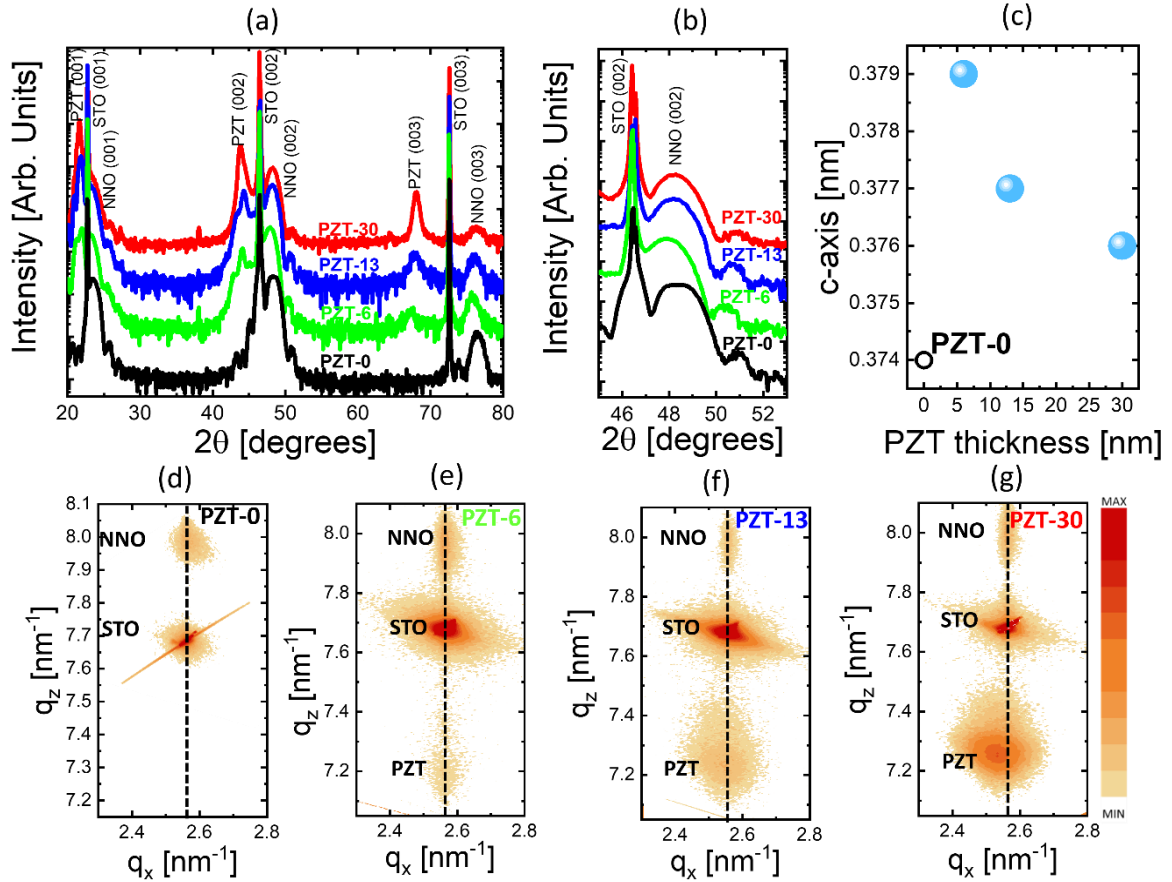

Figure 2 (a) XRD patterns for the PZT(d)/NNO(7 nm)//STO(001) sample series (d = 6, 13 and 30 nm) and STO(6 nm)/NNO//STO(001) with (b) related zoom around the NNO (002) diffraction peak that shows a clear trend of the c-axis parameter as quantified in (c). (d-g) Reciprocal space maps around the asymmetric STO (013) reflection.

We resorted to a Scanning Transmission Electron Microscopy (STEM) technique to get insights into the microscopic structural details. In particular, we measured the PZT-6 and PZT-30 thin films by STEM in the high-angular annular dark field (HAADF) imaging mode. The obtained images rendered and confirmed the island-type growth for the PZT-6 sample and a very continuous layer for the PZT-30 one as shown in Figures 3a and 3b, respectively. At higher resolution (Figs. 3c,d), the HAADF-STEM images indicated a homogenous downward polarization for the PZT-30 sample, and the coexistence of both polarizations for the PZT-6 one. For the latter, one can observe polarization domains of typical size comprised between 30 nm and 70 nm which are oriented either upwards (UP) either or downwards (DOWN). They are difficult to map, because, due to the TEM lamellae thickness, one might often observe in-depth superimposed domains. A lattice distortion of the PZT structure is clearly visible between two distinct polar domains, and the presence of such an inhomogeneous strain distribution in the PZT-6 sample can favour the existence of different polarization states via flexoelectricity. A lattice distortion of the PZT structure is clearly visible between two polar domains pointing in opposite directions, and the presence of such an inhomogeneous strain distribution in the PZT-6 sample can favour the existence of different polarization states. Indeed, flexoelectricity has already been reported to enable the reversal of self-polarization in thin films[32–35].

One can also notice that for both PZT-thicknesses (6 and 30 nm), the NNO layers present some Ruddlesden-popper (RP) spurious phases, which are commonly observed in perovskite nickelates as already reported previously for uncapped films [[36–38]]. RP phase could be a cause of the increase of the c-axis parameter of the layer, but their relative constant presence in different NNO layers does not allow to be conclusive on their influence on the c-axis of the NNO as a function of the PZT thickness. However, it can be observed that the spurious phases are mostly present in-between the islands/domains, which might indicate that the higher number of RP defects might be related to the island-growth mechanism. The strain distributions, and notably the out-of-plane lattice expansion $\varepsilon_{yy}$, have been investigated by geometric phase analysis (GPA)[39]. Figures 3e,f compare the out-of-plane lattice expansion across the NNO layers of PZT-30 and PZT-6 samples, respectively. The NNO layer of PZT-30 has an overall homogenous strain distribution, while the NNO layer of PZT-6 shows heterogeneities related to the island-growth. Furthermore, averaged strain profiles shown in Figure 3g indicate that NNO has a larger out-of-plane cell parameter in PZT-6 when compared to PZT-30, notably close to the interface with the PZT layer. The out-of-plane lattice constant is measured with respect to the $SrTiO_3$ substrate chosen as a reference. The NNO out-of-plane value is on average approximately 0.8% smaller for PZT-30 compared to PZT-6, in fair agreement with the XRD differences.

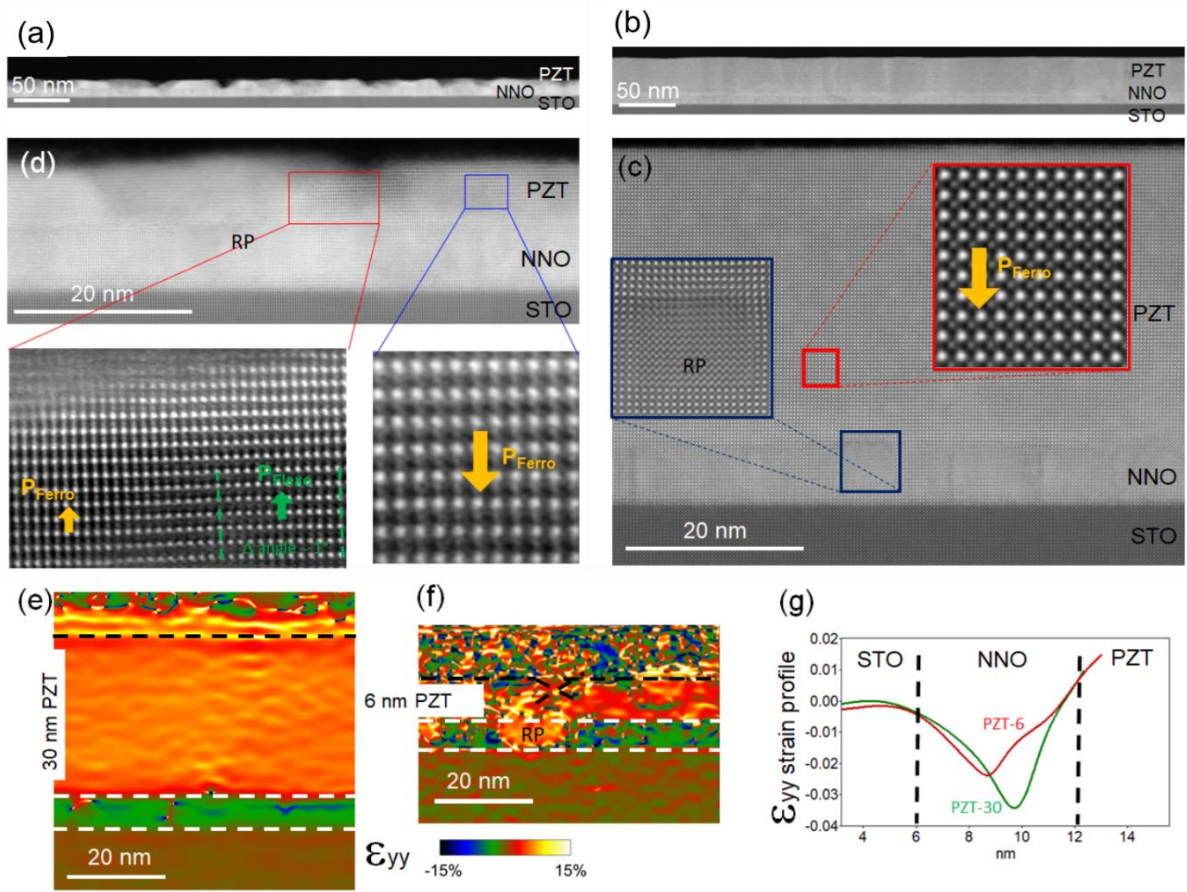

*Figure 3. (a-d) HAADF-STEM images of the (a,d) PZT-6 sample as observed along the [100]$_{pc}$ orientation, and (b,c) PZT-30 sample observed along the [110]$_{pc}$ crystal orientation. PZT-30 sample only shows a homogeneous downwards polar state. The PZT-6 sample shows more lattice distortions within the PZT layer. (d) Evidence of lattice distortions at the edge of a polar domain state, which may result in an upwards flexo-electric contribution. Due to the shape and distortion of each island/domain, the ferroelectric displacements are difficult to image, but*

*heterogeneities are obviously present and both upwards or downwards polarisations can be observed. (e,f) GPA maps of the out-of-plane expansion ($\varepsilon_{yy}$) measured with respect to the SrTiO$_3$ substrate taken as a reference, (g) averaged $\varepsilon_{yy}$ profiles.*

The phase diagram of bulk nickelates provides for NdNiO$_3$ concomitant onset temperatures for both MIT and AFM ordering (*i.e.* $T_{MIT} = T_{Néel}$). In the seminal paper of Catalan *et al.* the $T_{MIT}$ is identified with the peak observed in the d(ln($\rho$))/dT temperature-dependence[40], and used in some other works as main criteria[16,17,41]. However, also the minimum resistance which represents the zero-crossing point of the d(ln($\rho$))/dT temperature-dependence, is largely used[42–46]. It is clear that independently of the used criteria, to mark the onset of the MIT one has to assure that the expected bulk-like relationship with $T_{Néel}$ can be safely retrieved. Below we show that to recover the concomitant property $T_{MIT} = T_{Néel}$, one has to use the same criteria for both temperature-dependence of the transport and magnetic Bragg peak intensity. In this regard, we show in Figure 4a, for exemplative purposes, the resistivity curve of a NNO thin film grown onto a LaAlO$_3$ (LAO) substrate together with the temperature dependence of the (¼,¼,¼)$_{pc}$ magnetic Bragg peak acquired in a REXS experiment during a cooling process.

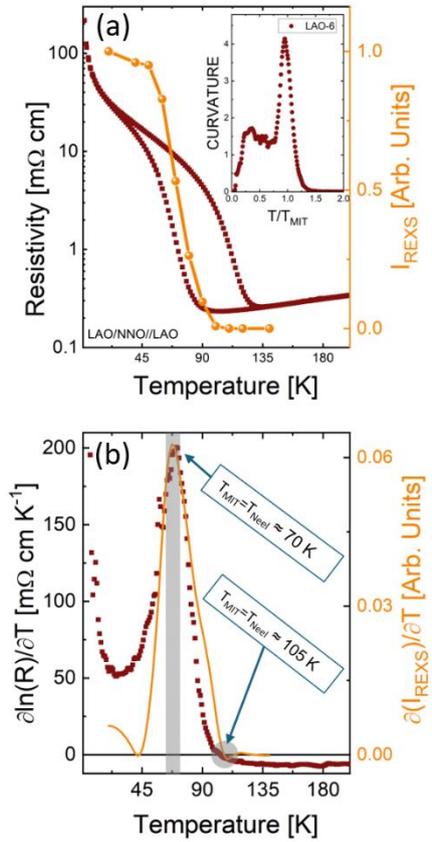

*Figure 4 (a) Resistivity (red curve) and REXS (orange curve) as a function of the temperature for LAO/NNO//LAO heterostructure. The inset show the temperature dependence of the curvature calculated for the cooling process. In (b) the derivatives of both curves are shown as explained in the text. The maximum and zero-crossing points are highlighted for both $T_{MIT}$ and $T_{Néel}$.*

Here, a better-matching substrate is used to assure that the expected bulk properties could be, indeed, easily reproduced for our NNO thin film. The temperature dependence of the resistivity shows a relatively sharp MIT with a large resistivity change accompanied by a large hysteresis, which points at a first order transition. The latter is mostly associated with the appearance of insulating (metallic) domains in a metallic (insulating) matrix in a cooling (warming) process, as already shown by several other techniques, such as PEEM[11,47] and C-AFM[11]. Figure 4b shows the d(ln($\rho$))/dT curve superimposed to the derivative of the AFM Bragg peak REXS intensity (d(I$_{REXS}$)/dT), as a function of the temperature. The latter was performed after the I$_{REXS}$(T) data set underwent an interpolation procedure. The expected $T_{MIT} \approx T_{Néel}$ relationship is retrieved when the same identification criteria is considered, *i.e.* peak position or zero-crossing point. The shadowed area of Figure 4b indicates roughly the temperature regions where both $T_{MIT}$ and $T_{Néel}$ can be identified as concomitant, hence, by using the same identification criteria one can study at which extent those values are different and, eventually, gather further information about the nature of the transition itself. In the following, both $T_{MIT}$ and $T_{Néel}$ are identified by using the peak position of the d(ln($\rho$))/dT and d(I$_{REXS}$)/dT temperature dependences, respectively. Finally, we would like to stress also the importance of the geometrical shape characterizing the $\rho$(T). Indeed, as will be clearer later, and especially in the case of nickelates, the shape of $\rho$(T) can depend upon fine details related to the electronic properties of the material itself and controlled by dimensionality and/or strain effects. We report in the inset of Figure 4a the curvature *c* calculated from the measured $\rho$(T) as a function of T/T$_{MIT}$ in a cooling process. We have used the following formulation where $\rho(T/T_{MIT})'$ and $\rho(T/T_{MIT})''$ are the first and second derivative of the rescaled resistivity [48]:

$$c = \frac{\rho(T/T_{MIT})''}{[1 + (\rho(T/T_{MIT})')^2]^{3/2}}$$

For the bulk-like expected s-like shape[40] of the $\rho(T)$, and reported in Figure 4a, it is worth noticing that the curvature presents at a temperature corresponding to the MIT a very sharp and intense peak, followed by a rather small one at lower temperatures. Substantial variations from this finding may speak in favor of modified electronic and magnetic properties of the nickelate thin films, which are indeed linked to a distinct metallic/insulating and AFM/paramagnetic domain textures. Once we have properly specified how both $T_{MIT}$ and $T_{Néel}$ will be identified, we show in Figures 5a-c the temperature dependence of the NNO resistivity as a function of the PZT thickness. As already specified above, the NNO layers have been grown with the same parameters and underwent the same oxygen annealing procedure, thus the calculated $T_{MIT}$ values mostly depend upon the properties of the PZT layer. Those values are indicated in each panel of Figure 5. Moreover, by a quick inspection, and by a direct comparison with the $\rho(T)$ reported in Figure 4a for the LAO-grown sample, one can note that the MIT of the entire sample series grown onto STO, including also the PZT-0 sample (Fig. 5d), exhibit largely modified shapes with smaller hysteresis. Those properties not only depend upon the different strain induced by the STO substrate (+2.5%), but also on the different PZT thicknesses and capping layer. The calculated curvatures, presented in the related insets of Figures 5a-d, show indeed clear differences. In particular, a unique and rather sharp peak is observed only in the case of the PZT-30 sample at $T_{MIT}$, and by decreasing the PZT thickness this peak progressively diminishes in intensity until fully disappears for the PZT-6 sample, as indicated by solid arrows in the corresponding insets. Finally, the peak at $T/T_{MIT} = 1$ for the curvature curve is well defined for the PZT-30 sample and largely comparable with the one obtained for the test sample PZT-0 as well as for the LAO-grown sample for which we have found that $T_{MIT} = T_{Néel}$. The presence of the curvature peak at $T_{MIT}$ suggests that the $\rho(T)$ shape inherits some information from the magnetic properties as well. On a simple phenomenological approach, we can suggest using the presence of this peak at $T/T_{MIT} = 1$ as a proxy of the concomitant properties of $T_{MIT}$ and $T_{Néel}$, however, to prove this fully a deeper study is necessary which is beyond the scope of this work.

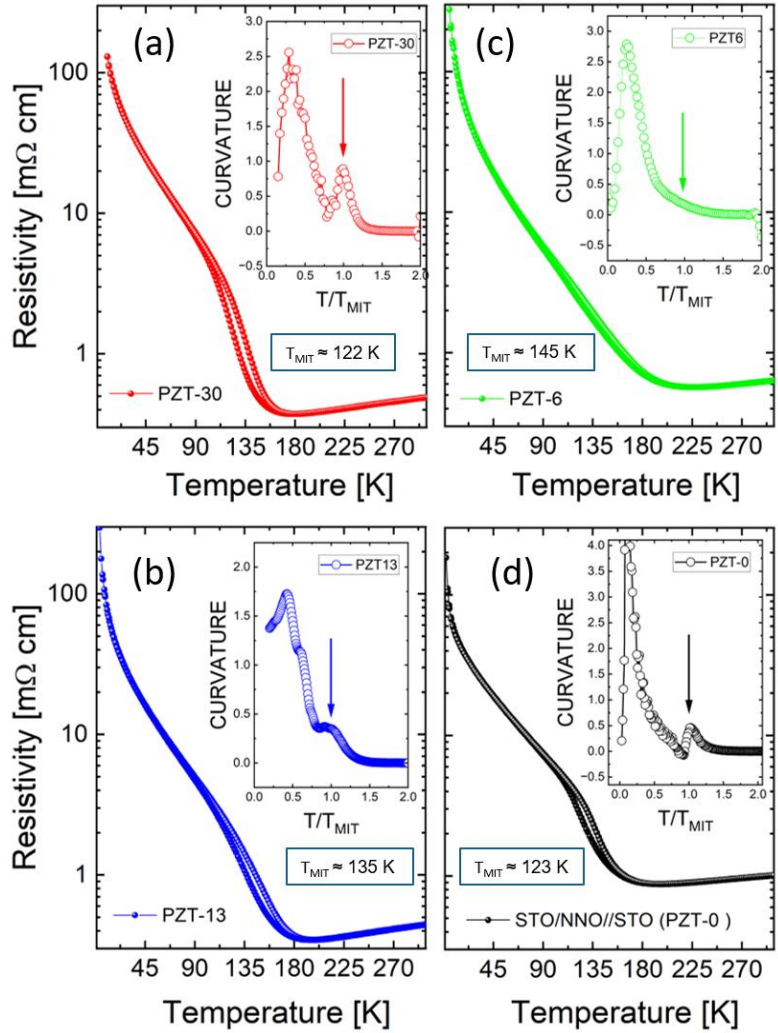

*Figure 5 (a-c) Resistivity as a function of the temperature for PZT/NNO//STO heterostructures with different PZT thicknesses (PZT-6, PZT-13 and PZT-30) and (d) a test sample PZT-0 STO(6 nm)/NNO//STO. The different $T_{MIT}$ are indicated in the specific insets of each panel. Curvature curves are sketched in the top inset and calculated for the cooling process.*

The observed PZT-induced changes in the onset temperatures for the MIT should therefore also correspond to a shift in the onset temperature for the AFM order, also considering the different intensity of the curvature peak at $T_{MIT}$. To retrieve the information about the magnetic transition we performed REXS measurements in the geometry sketched in Figure 6a (please refer to the Methods section for more details). The variation of the intensity of the magnetic Bragg peaks as a function of the temperature is shown in Figures 6b-d for the three different PZT thicknesses and acquired for a cooling process. The temperature evolution of the normalized REXS peak intensity for each sample is shown in Figure 6e, and the determined $T_{Néel}$ values are shown in Figure 6f together with the $T_{MIT}$ as a function of the PZT thickness. We found that for the lowest PZT thickness there is a very large discrepancy between the MIT and the AFM ordering onset temperatures with $T_{MIT} > T_{Néel}$. This discrepancy is progressively reduced upon increasing the PZT thickness until the bulk behavior[2] ($T_{MIT} = T_{Néel}$) is almost recovered in the case of the PZT-30 sample as, on the other hand, more clearly obtained for the test sample (PZT-0). We want stress here that, beyond the calculated values, this trend is observed also in the case the zero-crossing point is used as identification criteria. This demonstrates that the intrinsic properties of our sample series are completely independent from the analysis approach used to quantify both $T_{Néel}$ and $T_{MIT}$ quantities.

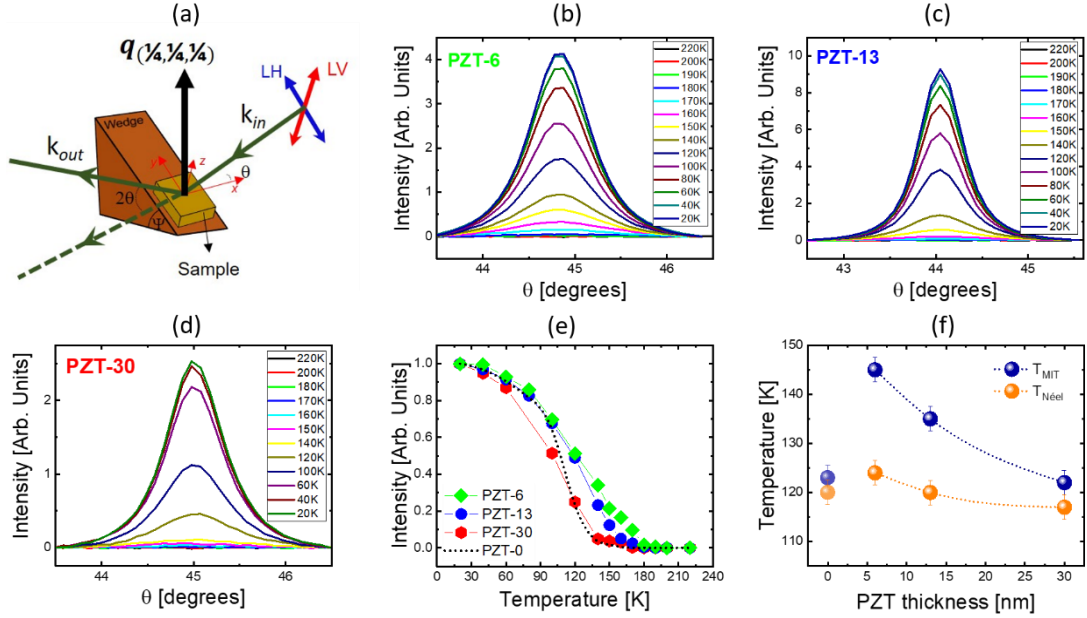

*Figure 6 (a) Measurement geometry with a 45°-tilted Cu-wedge. The [111] and [100] crystallographic axes define the scattering plane. Measurements performed with light polarized linear vertical (LV). Magnetic Bragg peak measured as a function of temperature in a cooling process for the (b) PZT-6 (c) PZT-13 and (d) PZT-30 samples. (e) Temperature evolution of the normalized maximum intensity of the magnetic reflection for the three different PZT thicknesses together with the test sample PZT-0. (f) PZT thickness dependence of both $T_{MIT}$ (Cyan) and $T_{Néel}$ (Orange) with dash lines used as eyes-guide.*

Atomic level spectroscopy elucidates further details on the structural peculiarities of the NNO films. We performed XAS measurements at both O K- and Ni $L_{3,2}$-edges in fluorescence yield mode and at 40 K in order to gather information about the electronic structure of our samples. Figure 7a shows the XAS O K-edge of PZT-6 and PZT-13 samples which exhibit a very comparable pre-peak feature at the O K edge (ca. 528 eV), as expected for a negative charge transfer insulator material, and corresponding to the transition from O 1s core level to the ligand hole L̲ ($3d^8$L̲ ground state). The pre-peak at the O K-edge is the same between the two samples which indicates that the two different PZT thickness does not affect the self-doped holes concentrations in the hybridized 3d bands. It therefore provides a further indication that no relevant variation of oxygens vacancies can be ascribed to the different PZT thickness. Indeed, in the presence of a large amount of oxygen vacancies the O K-edge pre-peak should disappear suggesting the filling of the ligand band. Moreover, according to Li *et al.*, the oxygen vacancies are found to largely modify the electronic configuration of the negative charge transfer nickelate thin films[49] and, more importantly, they do not alter the AFM ordering. In Figure 7b we show a zoom around the Ni-$L_3$ XAS features for the entire sample series together with the one acquired for the test sample (PZT-0). All the XAS spectra were characterized by a clear doublet-like feature, the components of which are here indicated as L and S, that correspond to a different electronic configuration as already introduced before and recalled in Figure 7b. The long (L) Ni-O bond[10], holds the largest magnetic signal (S=1), which is indeed confirmed by the photon-energy dependence of the magnetic Bragg intensity at $q_{AFM}$ = (¼,¼,¼)$_{pc}$ and shown in Figure 7b (black dotted line). One can observe that for the entire sample series the amplitude of the L-feature (*ca.* 853 eV) is higher than that of the S-feature (*ca.* 854.5 eV), in strong contrast with what is usually observed for NNO in both bulk and thin film form[9] (please refer to the PZT-0 black curve). We also observe that, when increasing the thickness of the PZT layer, the intensity of the L-peak steadily decreases at the expense of the S-peak one. The XAS differences are summarized in Figure 7c where the L/S intensity ratio is plotted. The thicker the PZT, the closer the L/S intensity ratio value is to the one obtained for the PZT-0 sample.

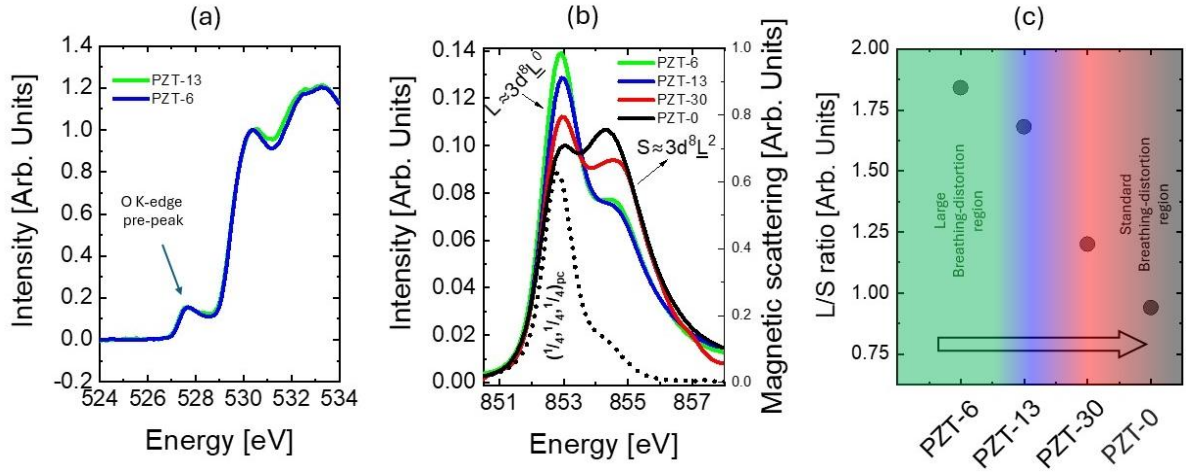

*Figure 7 (a) XAS spectra at the O-K edge for the PZT-6 and PZT-13 samples (b) XAS at the Ni $L_3$-edge for different PZT thicknesses at 40 K and in FY mode. All the spectra are area normalized. We also show for a complete comparison the XAS spectrum acquired for the test sample PZT-0 (black curve). The corresponding photon-energy of the magnetic scattering along the $q_{AFM}=(¼,¼,¼)_{pc}$ is also shown (dotted line). The L feature of the XAS has the maximum magnetic response. (c) L/S intensity ratio of the XAS spectra indicate a clear dependence upon the PZT thickness with a trend towards the value calculated for the test sample characterized by $T_{MIT}=T_{Néel}$.*

## Discussion and conclusion

Our study put in evidence how structural, magnetic and transport properties in negative charge transfer systems are intimately connected. We have shown that the NNO functional properties can largely depend upon the presence of a PZT top-layer characterized by different strain states and polarization textures. It is worth to recall here that the presence of oxygen vacancies and RP-like spurious phases are usually linked to an expansion of the c-axis in nickelate thin films caused by a decrease of the nominal $Ni^{3+}$ valence state[50–52]. From our 'local' STEM results, this possibility can be safely excluded which demonstrates that, indeed, the volume rich in RP-like structures is largely comparable between each sample. On the other side, more 'bulk-like' XAS measurements at the O K-edge presented in Figure 7a, fully exclude any possible role played by a different amount of oxygen vacancies in our samples. All those experimental findings clearly point at some peculiar lattice distortions[40] of the NNO unit cell. Moreover, the observed variation of the $T_{MIT}$ and $T_{Néel}$ values as a function of the PZT thickness points also at a modulation of both electronic and magnetic properties of the NNO thin films. Finally, summarizing all the experimental finding, we could observe that NNO layers grown adjacently to a fully strained or partially relaxed and multidomain PZT top-layer are characterized by a $T_{MIT}$ larger than $T_{Néel}$, while for a PZT top-layer fully relaxed and exhibiting a monodomain structure: the bulk-like $T_{MIT} = T_{Néel}$ relationship could be restored. Thus, a fully strained multidomain PZT layer leading to $T_{MIT} > T_{Néel}$, points at intrinsic NNO structural distortions as dictated by the bulk nickelate phase diagram. Some useful information can be gathered by the XAS spectra at the Ni $L_3$-edge which indeed show clear modulations of both L and S features as a function of the PZT thickness (Fig. 7c). The L/S intensity ratio holds a link to the strength of the breathing distortion in perovskite rare-earth nickelates as emphasized by the double cluster model put forward by Green et al.[10]. In particular, an enhancement of the breathing distortion mode is found for increasing L/S intensity ratio. It has been previously mentioned that these L and S peaks are stemming from the main contributions of the long ($d_L$) and short ($d_S$) Ni-O bonds, respectively, which exhibit specific energy resonance in the breathing mode structure, characterized as $d_{L/S} = d_0 \pm \delta d$, with $d_0$ being the mean value. Therefore, in a simple and qualitative approach, we can suggest that the NNO of the PZT-6 sample presents a larger breathing distortion than the one associated to the PZT-30 sample. This has to be reconducted to the peculiar structural properties of the fully strained and multidomain PZT layer. A fully strained PZT layer will apply a stronger structural alteration of the neighbor's $NiO_6$ octahedra at the PZT/NNO interface via polar distortions, therefore causing also the NNO c-axis expansion. Although, the role played by the oxygen vacancies about structural distortion and physical properties is largely documented in nickelates[51,53–55], all our experimental finding combined with the optimal growth conditions of the heterostructures delineate a clear situation where we can fully exclude their involvement. Finally, it is worth noticing that if oxygen vacancies play the major role for the reported interfacial structurally-driven distortion of the $NiO_6$ octahedra, these should be clearly independent from the PZT thickness. Finally, we have shown that it is possible to modulate the degree of lattice distortion in a nickelate thin film via a combination of polarization texture and strength of the strain state of the PZT top-layer. The modification of the XAS Ni $L_3$-edge features offers a direct link to the alteration of the breathing distortion pattern which can be explained as the direct consequence of the PZT-induced polar distortions which remain to be demonstrated. Indeed, further STEM investigations with the possibility to image also the oxygen positions are necessary to effectively prove that intrinsic polar distortions[29] propagating from the PZT towards the NNO unit cells, locally affect the Ni-O-Ni bond angle/length. Those results together with the ones already presented here, can open new avenues in the control over both $T_{MIT}$ and $T_{Néel}$ in nickelate thin films.

**Acknowledgments**

This work was funded by the French National Research Agency (ANR) through the ANR-21-CE08-0021-01 'ANR FOXIES' and, within the Interdisciplinary Thematic Institute QMat, as part of the ITI 2021 2028 program of the University of Strasbourg, CNRS and Inserm, it was supported by IdEx Unistra (ANR 10 IDEX 0002), and by SFRI STRAT'US project (ANR 20 SFRI 0012) and ANR-11-LABX-0058_NIE and ANR-17-EURE-0024 under the framework of the French Investments for the Future Program. The authors acknowledge the XRD platform of the IPCMS, and Jérôme Robert, Laurent Schlur and Gilles Versini for technical support.

*Email : daniele.preziosi@ipcms.unistra.fr

[⊥]Present address: Department of Applied Physics, Eindhoven University of Technology, 5600MB Eindhoven, The Netherlands. Mail: g.f.k.krieger@tue.nl